\documentclass[aps,twocolumn,showpacs,superscriptaddress]{revtex4-1}
\usepackage{amsmath,amssymb,graphicx,epstopdf,color,bm,soul}

\begin{document}

\title{Polariton condensation in photonic crystals with high molecular orientation}

\author{D. V. Karpov}
\affiliation{Institute of Photonics, University of Eastern Finland, P.O. Box 111 Joensuu, FI-80101 Finland}
\affiliation{ITMO University, St. Petersburg 197101, Russia}

\author{I. G. Savenko}
\affiliation{Center for Theoretical Physics of Complex Systems, Institute for Basic Science (IBS), Daejeon 34051, Republic of Korea}
\affiliation{Nonlinear Physics Centre, Research School of Physics and Engineering, The Australian National University, Canberra ACT 2601, Australia}

\begin{abstract}
We study Frenkel exciton-polariton Bose-Einstein condensation in a two-dimensional defect-free triangular photonic crystal with an organic semiconductor active medium containing bound excitons with dipole moments oriented perpendicular to the layers. We find photonic Bloch modes of the structure and consider their strong coupling regime with the excitonic component. Using the Gross-Pitaevskii equation for exciton polaritons and the Boltzmann equation for the external exciton reservoir, we demonstrate the formation of condensate at the points in reciprocal space where photon group velocity equals zero. Further, we demonstrate condensation at non-zero momentum states for TM-polarized photons in the case of a system with incoherent pumping, and show that the condensation threshold varies for different points in the reciprocal space, controlled by detuning.

\end{abstract}

\pacs{78.67.Pt,78.66.Fd,78.45.+h}
\maketitle

%--------------------------------------------------------------

%{\it Introduction.---} 
\section{Introduction}
The large exciton binding energy and oscillator strength of organic materials embedded in light-confining structures such as optical cavities make it possible to achieve giant energies of the Rabi oscillations desired for room-temperature exciton-polariton (EP) condensation~\cite{Daska2014, Plumhof2013, Crist2007}. In this respect, two-dimensional (2D) photonic crystals (PC), which can be easily integrated with organic materials~\cite{Kaname, Yamao}, are a current area of focus. Low group velocity of the optical Bloch modes at the edge of the conduction band provides for a long lifetime of the slow waves and thus seems promising for the realization of polariton condensation, similar to the enhancement of coherent emission in defect-free photonic crystals~\cite{NojimaBasic,NojimaPRB,Nojima2001,Notomi}.

A considerable number of organic semiconductors, such as thiophene/phenylene co-oligomer single crystal, 1,4-bis(5-phenylthiophen-2-yl) benzene and 2,5-bis(4-biphenyl)thiophene~\cite{Kaname, Yamao}, have transition dipole moments oriented along the vertical direction with respect to the main crystal face. For this reason, these organic crystals are inappropriate for strong interaction with the optical modes of a Fabry-Perot cavity, where the electric field is oriented perpendicular to the dipole moment. Instead, as we will show in the case of 2D PCs, transverse magnetic (TM) modes have the electric field component perpendicular to the plane of the crystal and therefore can be strongly coupled with excitons. It can be noted that there exist other materials, such as cyano-substituted compound 2,5-bis(cyano biphenyl-4-yl) thiophene, in which the transition dipole moment lies in the in-plane direction with respect to the crystal face. While such materials can be assumed to demonstrate strong coupling with the Fabry-Perot cavities or transverse electric (TE) modes of PCs~\cite{PRX}, supporting $\Gamma$-point condensation in the reciprocal space, this case is trivial and beyond the scope of our manuscript.

In this manuscript, we consider a 2D PC represented by a triangular lattice of pillars supporting the emergence of band gaps for both TE and TM polarizations. In principle, 2D PCs provide two types of exciton-photon quasiparticles. 
The first type results from coupling between excitonic and photonic modes below the light cone (free photon dispersion), with such modes called guided PC polaritons. 
The second type, called radiative polaritons, constitutes the excitons lying above the light cone. Polaritons of the latter type can be effectively analyzed by angle-resolved spectroscopy, under the condition that the exciton-photon coupling is much greater than the intrinsic photon linewidth~\cite{GeraceExp}. These two modes can be employed differently; in particular, the radiative modes can be used as efficient reflectors~\cite{reflect}, whereas the guided modes can be utilized for the realization of strong light-matter interaction in such devices as vertical cavity surface emitting lasers~\cite{vecs} and polariton lasers~\cite{GeraceTheory}.

Both types of polaritons can be made of photons representing slow Bloch modes (SBM), confined in 2D PCs in the vicinity of the extremum points at the edge of the photonic band gap, where the group velocity approximately equals zero~\cite{NojimaBasic}. Slow velocity of the modes results in long optical paths, or in other words, nearly total light confinement. This effect has been used for mode synchronization in organic lasers~\cite{Notomi}, and has also proved to be efficient for lasing threshold reduction in vertical cavity surface emitting lasers~\cite{CosendeyApl} and in 2D PC lasers in the strong coupling regime. Notably, such lasers exhibit much lower threshold gains~\cite{NojimaPRB}.

\begin{figure}[t!]
\includegraphics[width=0.8\linewidth]{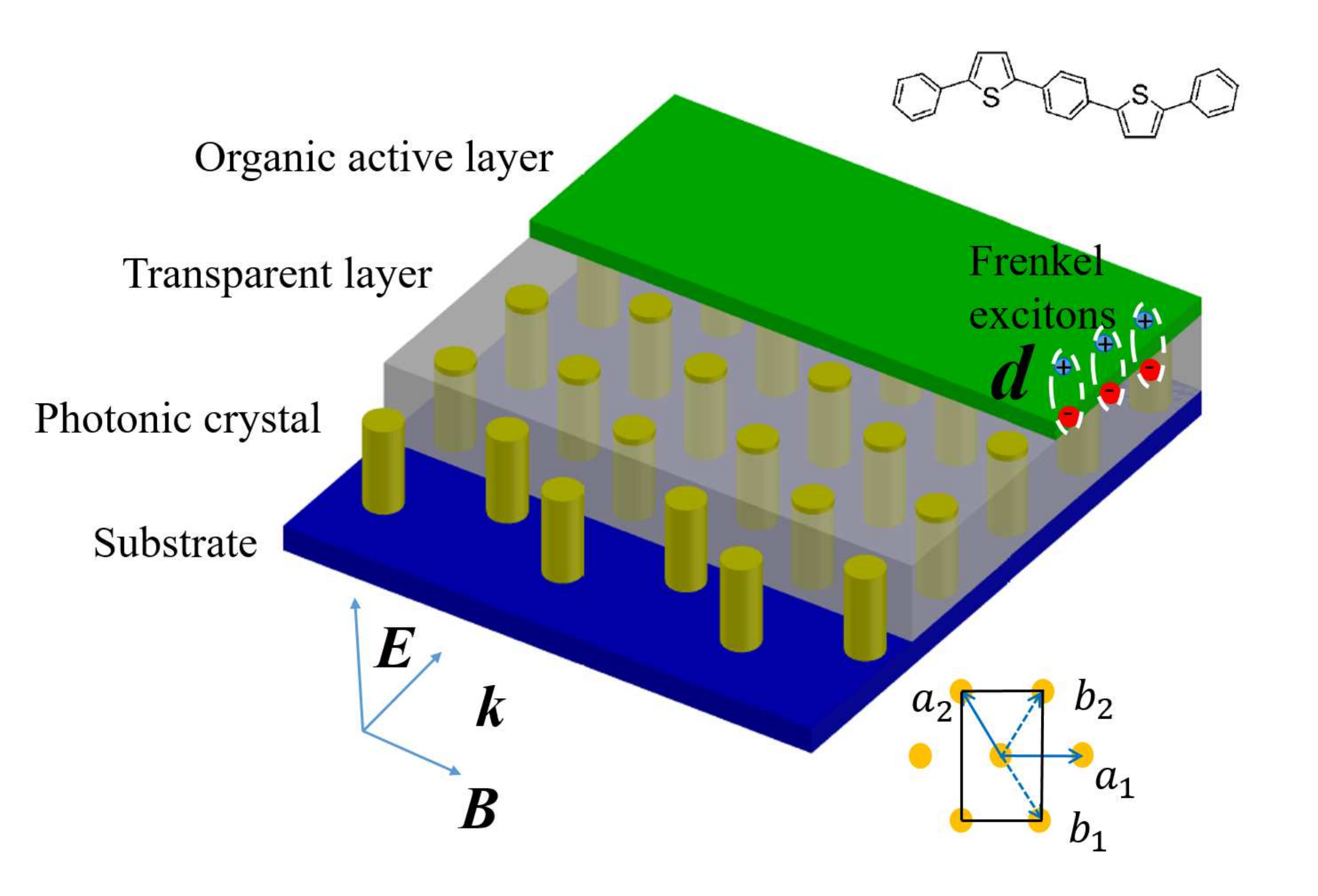}
\caption{(color online) Hexagonal photonic crystal with lattice vectors $a_1$, $a_2$ and reciprocal lattice vectors $b_1$, $b_2$ covered by a transparent layer of a polymer (semi-transparent grey) and a high molecular orientation organic active layer 1,4-bis(5-phenyl thiophen-2-yl)benzene (green). The system can be excited by an electromagnetic field through the edges.}
\label{fig:Fig1-eps-converted-to}
\end{figure}

The lifetime of the Bloch modes is mostly determined by the lateral quality factor of the PC, which itself depends on the size and quality of the nanostructure. The vertical quality factor can be considered infinitely high for 2D PCs of sizes greater than a hundred microns.

%--------------------------------------------------------------

%{\it Photonic band structure ---}
\section{Photonic band structure}
Our system schematic is presented in Fig.~\ref{Fig1}. The structure consists of aluminum nitride (AlN) pillars of radius 450 nm forming the photonic crystal, with a lattice constant of 1 $\mu$m and a refractive index ($n$) of 2.15. The transparent layer consists of a polymer material with optical properties close to air. The substrate provides for optical confinement in the vertical direction and should therefore be chosen from materials with refractive indices higher than that of AlN (e.g. GaN). However, we consider the system to be effectively 2D in our calculations and neglect the influence of the substrate on the optical properties of the system. Further, in order to make the light-matter coupling effective and strong, the thin active layer is firmly set onto the PC.  
The structure is exposed to an electromagnetic wave falling at the edge of the PC, nearly parallel to the layers of the structure. It should be noted that normal incidence does not provide for light confinement, which is necessary for a finite photon lifetime in the system, and is thus outside the focus of our study. 

We employ a standard Fourier method in order to decouple the Maxwell's equations and find independent behavior of the TE and TM polarizations, using separate equations for the magnetic and electric fields. Both the magnetic and electric fields are coplanar to the $z$ axis. In the case of TE polarization, we can represent the fields in the following form~\cite{Plihal,PlihalHex}:
\begin{eqnarray}
\label{hte}
\mathbf{H(\mathbf{r})}&=&(0,0,H_z(\mathbf{r})),\\
\label{ete}
\mathbf{E(\mathbf{r})}&=&(E_x(\mathbf{r}),E_y(\mathbf{r}),0),
\end{eqnarray} 
where $\mathbf{r}(x,y)$ is the coordinate in the $xy$ plane. 
Inserting \eqref{hte}~and~\eqref{ete} into Maxwell's equations in the frequency domain, we find:
\begin{eqnarray}
\label{hfield1}
\frac{\partial}{\partial x} \frac{1}{\epsilon(\mathbf{r})} \frac{\partial H_z(\mathbf{r})}{\partial x} +\frac{\partial}{\partial y} \frac{1}{\epsilon(\mathbf{r})} \frac{\partial H_z(\mathbf{r})}{\partial y} +\frac{\omega^2}{c^2} H_z(\mathbf{r})=0.
\end{eqnarray}
Further, we use the Bloch decomposition,
\begin{eqnarray}
\label{fourierH}
H_z(\mathbf{r})=\sum\limits_{G} A_G(k) e^{-i(\mathbf{k}\cdot\mathbf{r}+ \mathbf{G}\cdot\mathbf{r})},
\end{eqnarray}
where the summation is taken over the reciprocal lattice vectors, $G=n\bold{b_1}+m\bold{b_2}$, where $n$ and $m$ are integer numbers. Inserting~\eqref{fourierH} into \eqref{hfield1}, we obtain the eigenvalue problem for the TE-modes:
\begin{eqnarray}
\label{eigenTE}
\sum\limits_{G'} \epsilon_{G,G'}^{-1} (\mathbf{k}+\mathbf{G'})\cdot (\mathbf{k}+\mathbf{G}) \mathbf{A}_G=\frac{\omega^2}{c^2} \mathbf{A}_{G},
\end{eqnarray}
where $\epsilon_{G,G'}$ is the dielectric permittivity.

Let us now consider TM polarization, where
\begin{eqnarray}
\label{efield45}
\mathbf{E(\mathbf{r})}&=&(0,0,E_z(\mathbf{r})),\\
\label{efield42}
\mathbf{H(\mathbf{r})}&=&(H_x(\mathbf{r}),H_y(\mathbf{r}),0).
\end{eqnarray}
Then we write down Maxwell's equation,
\begin{eqnarray}
\label{hfield}
\frac{1}{\epsilon(\mathbf{r})} \frac{\partial^2 E_z(\mathbf{r})}{\partial^2 x} +\frac{1}{\epsilon(\mathbf{r})} \frac{\partial^2 E_z(\mathbf{r})}{\partial^2 y} +\frac{\omega^2}{c^2} E_z(\mathbf{r})=0,
\end{eqnarray}
and by substituting the ansatz
\begin{eqnarray}
\label{fourierTE}
E_z(\mathbf{r})=\sum\limits_{G} B_G(k) e^{-i(\mathbf{k}\cdot\mathbf{r}+ \mathbf{G}\cdot\mathbf{r})}
\end{eqnarray}
we arrive at the eigenvalue problem:
\begin{eqnarray}
\label{EqEigen}
\sum\limits_{G'} \epsilon_{G,G'}^{-1} (\mathbf{k}+\mathbf{G'})^2 \mathbf{B}_G=\frac{\omega^2}{c^2} \mathbf{B}_{G}.
\end{eqnarray}

It should be noted that the Fourier approach employed here is only valid for perfectly periodic photonic crystals. In reality, one should also account for the fact that the 2D PC is not infinitely large or lossless. For dielectric materials with small absorption coefficients, one can apply an effective Fourier approach in which the imaginary part of permittivity is considered as a perturbation~\cite{NojimaBasic}. In order to account for the finite lifetime of photons, we use experimental data for the absorption coefficient~\cite{aln} which gives us an imaginary part of the permittivity of about $10^{-4}$. Then we use the complex-valued permittivity and find complex eigen frequencies which give us the photon lifetime. The finite size of PCs and imperfections of a real sample lead to a reduction in photon lifetime, the effect of which is important for the Bloch waves under the light cone. For these modes, material losses do not reduce the photon lifetime, and imperfections of the geometry can be considered as the only source of dissipation. We include this mechanism of energy relaxation in our model phenomenologically, as discussed below.

\begin{figure}[!t]
	\includegraphics[width=0.55\linewidth]{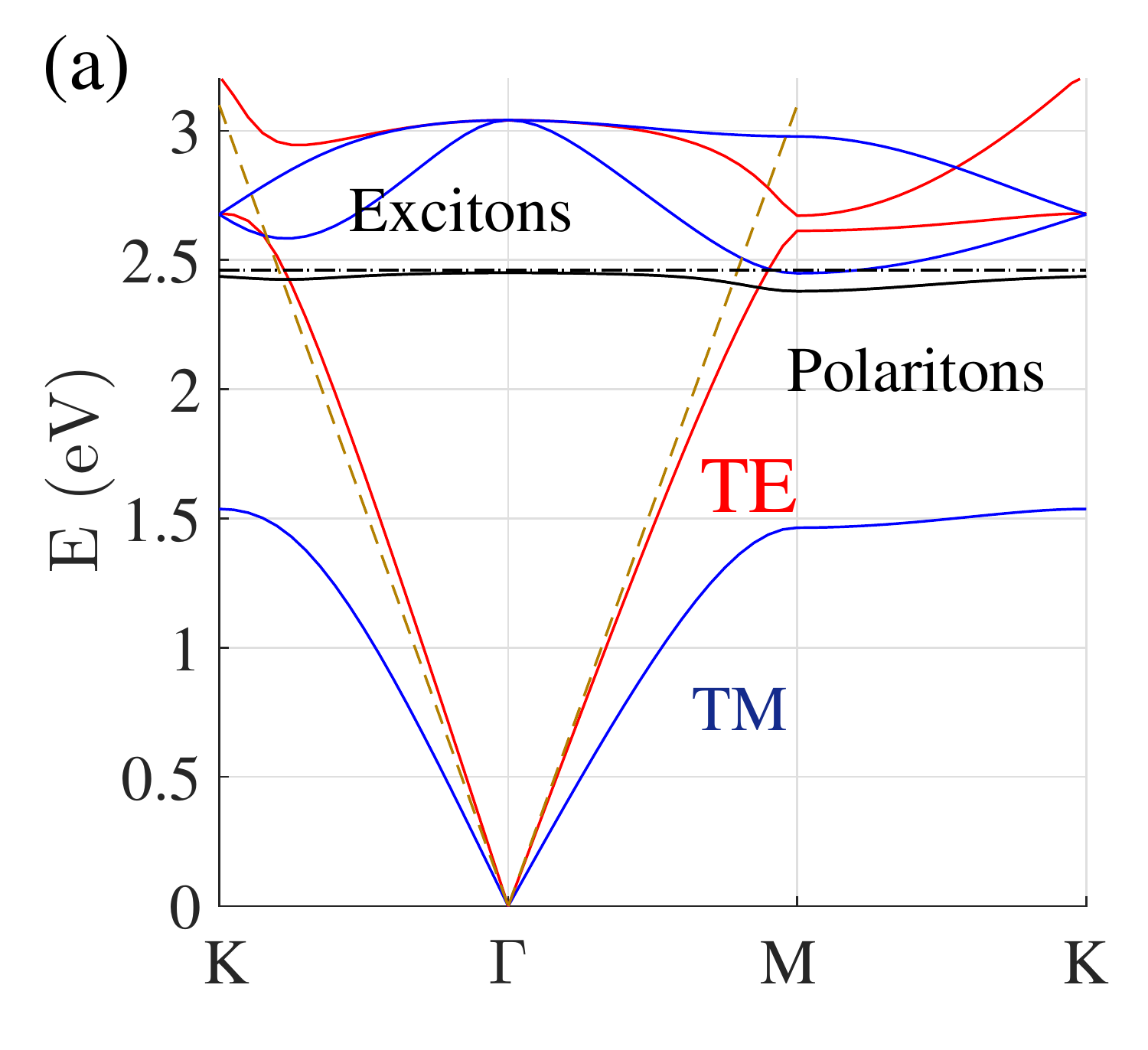}
	\includegraphics[width=0.42\linewidth]{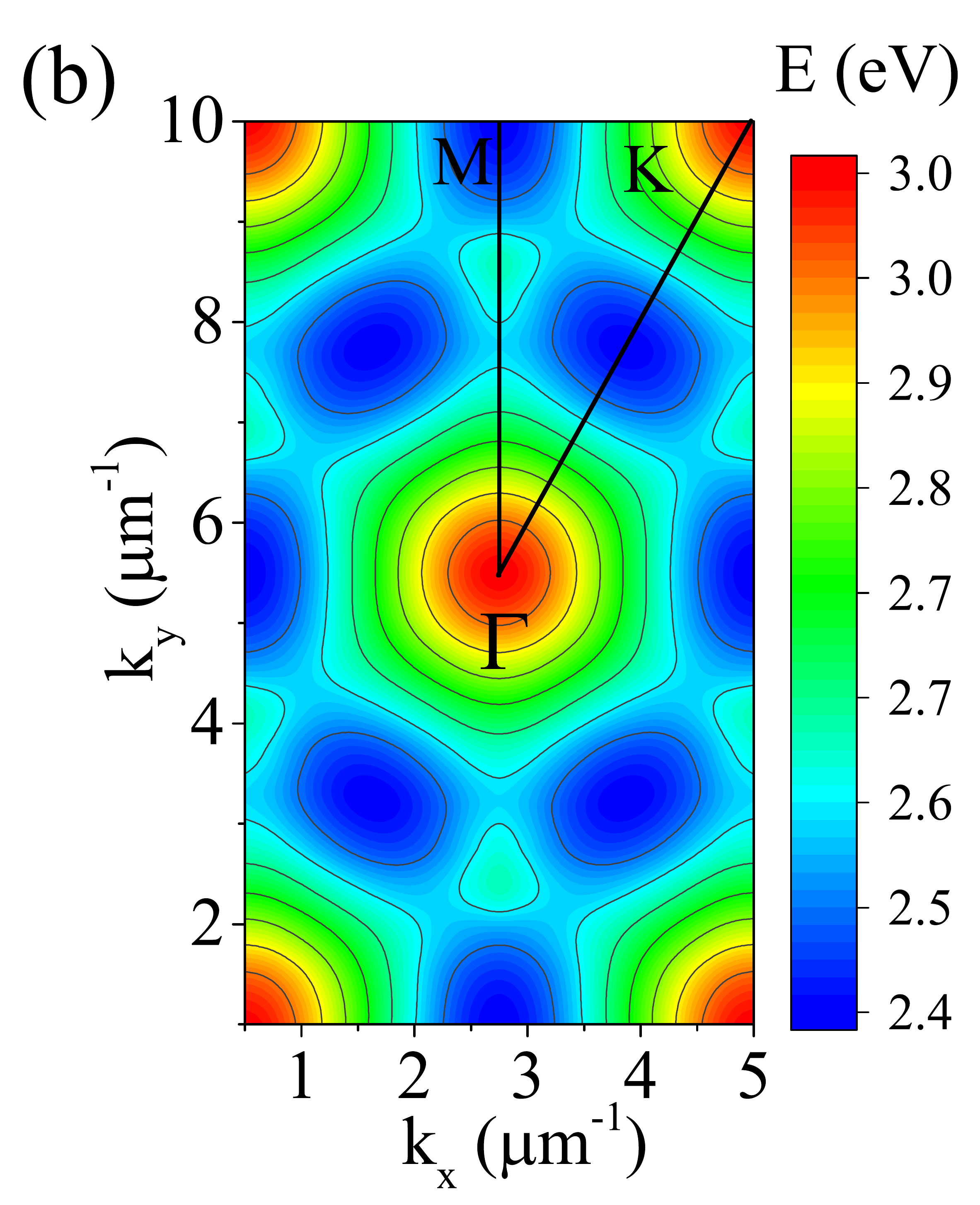}
	\caption{(color online) (a) Band structure of the 2D photonic crystal (presented in Fig.~\ref{Fig1}) showing TE-modes (red solid lines) and TM-modes (blue solid lines). The light cone and exciton dispersion are shown with the brown and black dashed lines, respectively. The lower polariton branch corresponds to the black solid curve. (b) A 2D photon dispersion for the TM-mode which we couple to excitons.} 
		\label{Fig2}
\end{figure}

When we apply Eq.~\eqref{EqEigen} to our geometry (Fig.~\ref{Fig1}), we find that the structure favors two energy minima in the spectrum of the TM modes (see Fig.~\ref{Fig2}). It is important that the TM mode supports the emergence of the SBM often associated with the emergence of the Van Hove singularity~\cite{VanHove}. One of the minima is located at the high-symmetry $M$-point of the Brillouin zone, and the other lies above the light cone in the vicinity of the $K$-point. Such modes (lying above the light cone) are usually referred to as radiated or quasi-guided modes since they can radiate in free space. The modes lying below the light cone (guided modes) are confined and can only radiate through imperfect sidewalls of the crystal or other disorders in the PC structure~\cite{GeraceTheory}.

%
%
%

%--------------------------------------------------------------

%{\it Quality factor ---}
\section{Quality factor}
The main condition for the strong coupling regime, which provides for EP formation, can be roughly stated as $\Omega_R>1/\tau_C,1/\tau_X$, where $ \Omega_R$ is the Rabi frequency, standing for the rate of energy exchange between the excitonic and photonic components, and $\tau_C$ and $\tau_X$ are the lifetimes of the photons and excitons, respectively. In case of organic polaritons, based on Frenkel excitons with typically long lifetimes, the lifetime of the particles is mostly determined by the photonic component. By definition, the latter is determined by the quality factor of the PC and frequency $\tau_C=2\pi Q/\omega_{real}$. The relation between the optical pumping rate and EP inverse lifetime determines the possibility of Bose-Einstein condensate (BEC) formation. Clearly, an increase of both the lifetime and the pumping rate allows one to achieve critical polariton concentration for BEC formation.

In 2D PCs, the full quality factor $Q$ can be found as the inversed sum of vertical and lateral quality factors: $Q^{-1}=Q^{-1}_{v}+Q^{-1}_{l}$. The lifetime of the Bloch modes, which lie above the light cone and can be coupled with free-space modes, is determined by the vertical quality factor which is a characteristic of the radiation losses in the perpendicular direction~\cite{Ferrier},   $Q^{-1}_{v}=-\omega_{real}/2\omega_{im}$, where  $\omega_{real}$, $\omega_{im}$ are the real and imaginary parts of the frequency of the Bloch modes~\cite{Sauvan2005}. We can estimate $Q_v$ of a 2D PC of final size $L$ using the assumption that such a PC supports modes with mean in-plane $k\propto \frac{1}{L}$. It is known that if, approximately, $L>100$ $\mu$m, then $Q_{v}> 50000$ \cite{Ferrier}. Thus the total $Q$ is mainly determined by the lateral losses.

In turn, $Q_l$ depends on the band structure, size, and disorder of the nanostructure~\cite{Sauvan2005},
\begin{eqnarray}
\label{qfac}
Q_{l}=\frac{\pi}{1-R(\lambda_0)} \left[ \frac{2cL^2}{\lambda_0 \alpha}\frac{1}{p\pi-\phi_r}-\frac{\lambda_0}{\pi} \left.\frac{d\phi_r}{d\lambda}\right|_{\lambda_0}\right],
\end{eqnarray}
where $R$ is the modal reflectivity, $\alpha$ is band curvature in the vicinity of the minimum (the second derivative of the dispersion), the group velocity can be expressed as $v_g=\alpha k$, $p$ is an integer number, and $\phi_r$ is the phase of the modal reflectivity at the edges of the 2D PC. In our simulations, we choose a typical value for the lateral quality factor: $Q_{l}\approx 2000$~\cite{Ferrier}. Polariton lifetime in this structure can be comparable with a Fabry-Perot microcavity at $\tau_p\approx  5$~ps.

Note that a typical quality factor of AlN hexagonal crystals is high enough for strong coupling, with such structures commonly used as microwire waveguides~\cite{appl1,appl2}. It is, however, insufficient to provide a thermal state for BEC, and therefore we will consider nonequilibrium condensation. Thus, our system is described with a kinetics approach.
Another beneficial property of AlN is that it has a small lattice mismatch with other nitride-based semiconductor alloys~\cite{ourAPL2016}; moreover, AlN stress-free layers can be easily grown on Si/SiC substrates~\cite{Kukushkin2016}.

%--------------------------------------------------------------

%{\it Dispersion relation ---} 
\section{Dispersion relation}
In organic active media, one can observe tightly bound Frenkel excitons with a typical size of one angstrom. This fact allows us to neglect the influence of the periodic structure of the 2D PC on the exciton wave function and consider non-uniform electric fields only. EPs emerge as mixed modes of the electromagnetic field and the exciton resonance, thus having the dispersion
\begin{eqnarray}
\label{disp}
\omega_{LP}(k)=\frac{\omega^{C}+\omega_k^{X}}{2}-
\frac{\sqrt{(\omega_k^{C}-\omega^{X})^2+\Omega_R^2}}{2},
\end{eqnarray}
where \cite{Kaname}:
\begin{eqnarray}
\label{Rabi}
\hbar\Omega_R=\sqrt{\frac{2|\mu|^2\hbar\omega_C(N/V)}{\epsilon}}.
\end{eqnarray}
The molecular packing density here can be estimated as $N/V\approx 10^{-3} \AA^{-3}$, $|\mu|\approx 25$~Debye. The estimations with formula~\eqref{Rabi} give exaggerated values of more than $1$ eV which do not comply with experimental data. This discrepancy is due to 99$\%$ of the excitons being uncoupled. For our simulations, we choose a value of Rabi energy typical for a planar geometry with an organic active layer, $\hbar\Omega_R\approx 100$~meV \cite{BP2T,Kaname}.  

\begin{figure*}[t!]
	\includegraphics[width=1.0\linewidth]{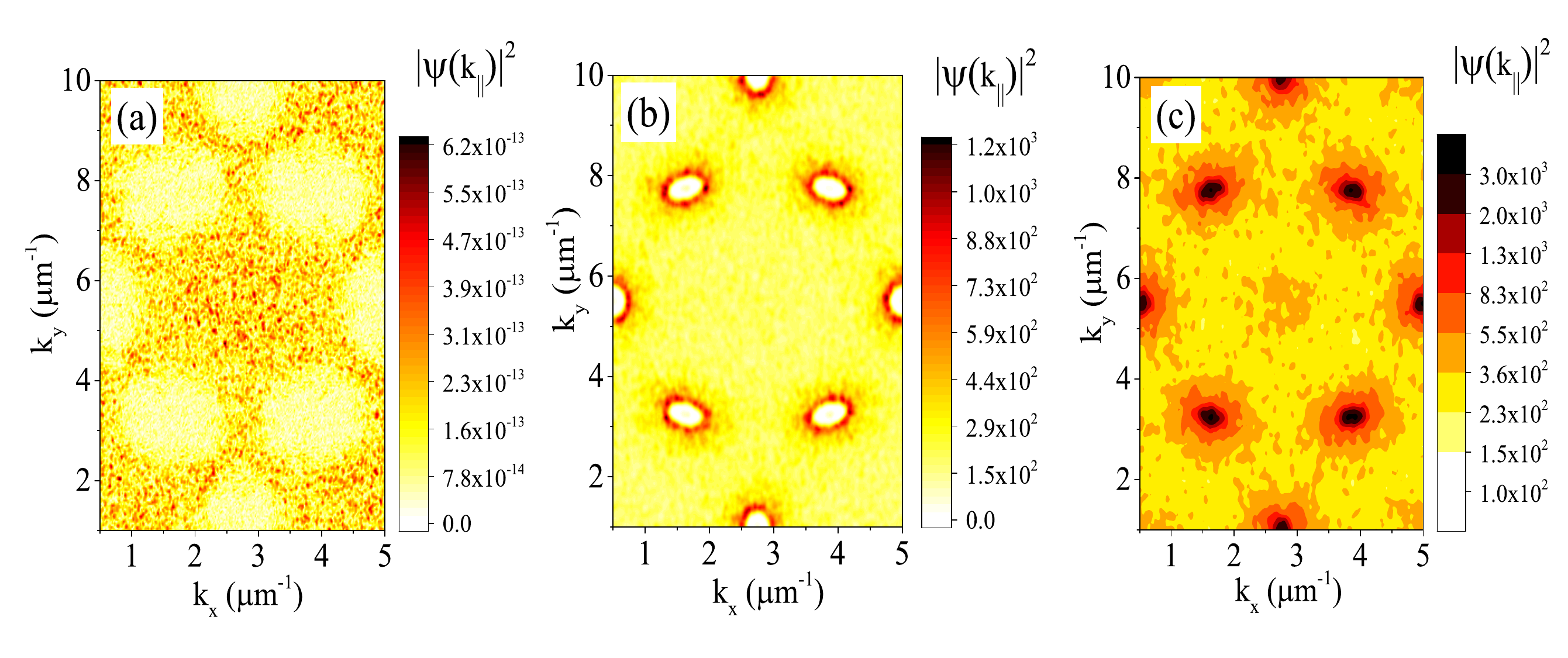}
	\caption{(color online) Colormaps of the exciton-polariton distribution in reciprocal space for different pumping powers: (a) $P= 0.01~ \mu m^2 s^{-1}$, below threshold thermal distribution; (b) $P=0.1~ \mu m^2 s^{-1}$, close to the threshold distribution, or bottleneck region; and (c) $P=0.3~ \mu m^2 s^{-1}$, above threshold.} 
	\label{Fig3}
\end{figure*}

While the Rabi energy in typical Fabri-Perot cavities with organic active regions varies between 100 and 800 meV~\cite{Org1,Org2,Org3}, in our case it takes the lowest limit since we have laminated the active layer on top of the PC. This configuration does not allow the achievement of full overlap between the electric field and the dipole. 
Such values of $\Omega_R$ typical for organic microcavities are much higher compared with GaAs/InGaAs quantum well-based microcavities due to the extremely high transition dipole moment that is typical for organic materials~\cite{Org1, Org2, Org3}. 
For a defect-free PC the electric field localizes in the Fabry-Perot microcavity, where the quantum well plays the role of a defect in 1D PC. But in the case of 2D PC, the active region covers the whole surface of the sample; therefore, the overcrossing of the dipole and the photon field in each elementary cell is small, yet for the whole sample the Rabi constant is quite high.

%PC quality factor due to material and geometrical losses does not provide electric field confinement in the active area, hence this fact does not allow such high values of strong coupling as could be possible for Fabri-Perot microcavities. It does allow though nonequilibrium condensation that can be described with a kinetics approach.

Figure~\ref{Fig2}a shows the results of the photonic band gap calculation: the red and blue lines correspond to TE and TM modes, respectively, and the black line shows the EP dispersion. 
The latter curve has two minima which can be considered as traps for polaritons, where their condensation might take place. 
It should be noted that we do not describe the polaritons based on TE-modes since first, most organic active regions have dipole moments oriented perpendicular to the surface of lamination, and second, the bandgap for TE-modes is less than the trap for the polaritons, resulting in nonzero group velocity and instability of the condensate.

\begin{figure}[!b]
	\includegraphics[width=1.0\linewidth]{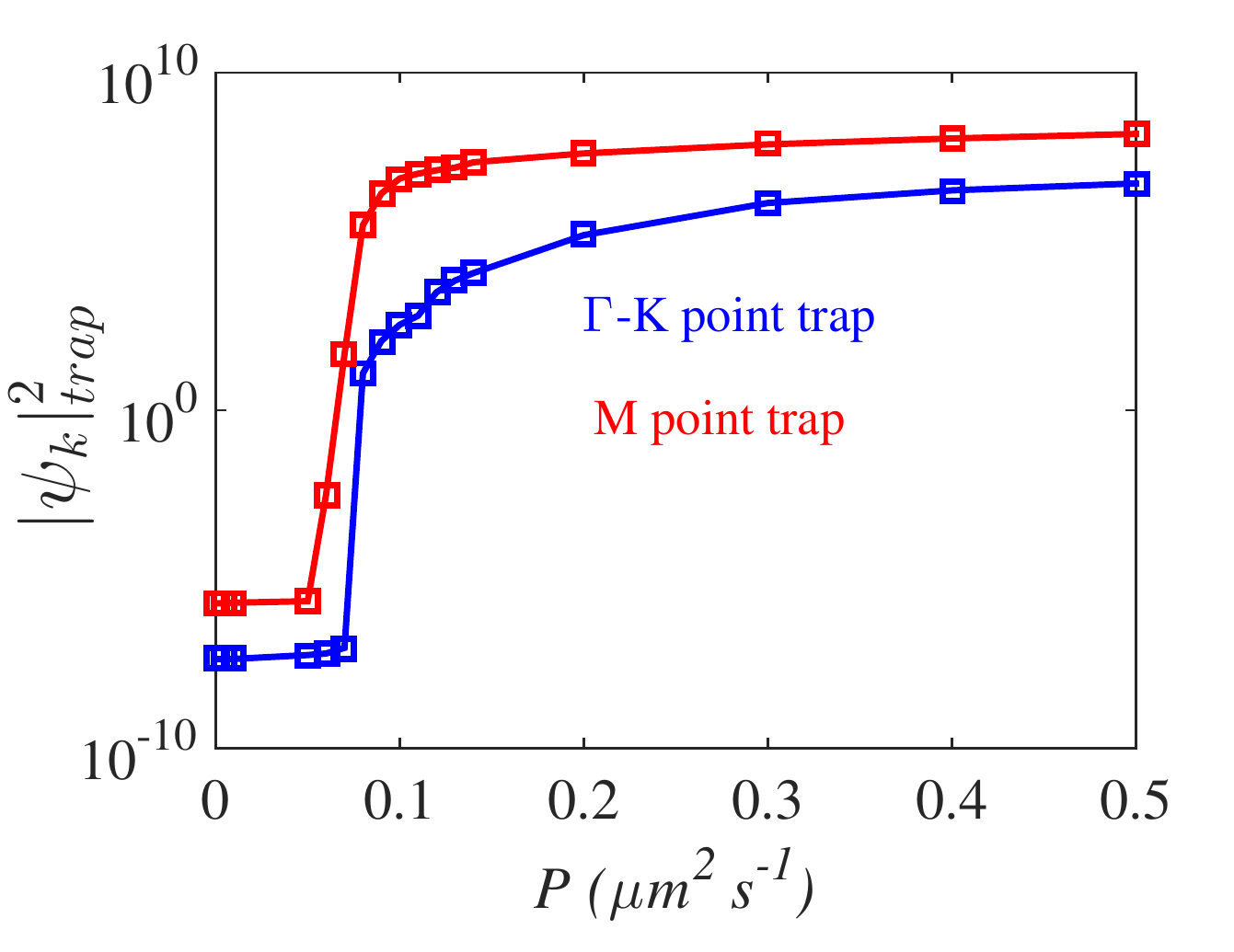}
	\caption{(color online) Exciton-polariton density in $ \mu m^{-2}$ at the bottom of the trap as a function of reservoir pumping.} 
	\label{Fig4}
\end{figure}
%
%
%

%-----------------------------------------------------------------------------------

%{\it Condensation kinetics ---}  
\section{Condensation kinetics}
After finding the bare EP dispersion, we can describe EP dynamics within the mean field approximation, where the EP field operator, $\hat{\Psi}(\mathbf{r},t)$, is averaged over the $z$-direction and treated as the classical variable $\psi(\mathbf{r},t)$ with Fourier image $\psi(\mathbf{k},t)$. The corresponding equation of motion reads~\cite{Wouters}:
\begin{align}
\label{eq:dpsixdt}
i\hbar\frac{d\psi(\mathbf{r},t)}{dt}&={\cal F}^{-1}\left[\hbar \omega_{LP}(k)\psi(\mathbf{k},t) -\frac{i\hbar}{2\tau(k)}\psi(\mathbf{k},t)\right]
\\
&+i\frac{\hbar\gamma}{2} n_X(\mathbf{r},t)\psi(\mathbf{r},t)+\alpha\left|\psi(\mathbf{r},t)\right|^2
\psi(\mathbf{r},t),
\notag
%\right.
&\hspace{0mm}
\end{align}
where ${\cal F}^{-1}$ is the inverse Fourier transform, $\alpha$ is a parameter describing the strength of particle-particle interactions, and $\tau$ is polariton lifetime. It can be estimated as: $\alpha\approx (10^{-22}/L)~ eV$ $cm^3$, where $L$ is the thickness of the active layer~\cite{Daska2014}. We use a value which is three orders of magnitude less than what we usually have in GaAs microcavities~\cite{bottleneck}. On one hand, such a small value of $\alpha$ does not lead to a significant blueshift. On the other hand, the main driver of condensation is still the cubic term in Eq.~\eqref{eq:dpsixdt}. The term $-i(\hbar/2\tau)\psi$  accounts for the radiative decay of particles. Safely assuming that exciton lifetime $\tau_X$ lies in the nanosecond range and is much greater than photon lifetime, we consider that polariton lifetime $\tau$ is determined mostly by the microcavity photon lifetime, which is $\tau_C=1/Im[\omega^C_k]$ where $\tau_C$ is determined by the material properties and geometry of the PC (see the discussion above).

Now we model the evolution of the exciton density in the reservoir, $n_X$, using the equation~\cite{Wouters}:
\begin{eqnarray}
\label{EqExcitons}
\frac{\partial n_X(\mathbf{r},t)}{\partial t}=P-\frac{n_X}{\tau_X}-\gamma~ n_X|\psi(\mathbf{r},t)|^2,
\end{eqnarray}
where $\tau_X$ is exciton lifetime, $P$ is the incoherent pumping power, and $\gamma$ is the rate of polariton formation fed by the excitonic reservoir.

%--------------------------------------------------------------
%{\it Results and discussion.---}
Using~\eqref{eq:dpsixdt} and~\eqref{EqExcitons} we calculate polariton distribution in the reciprocal space (Fig.~\ref{Fig3}). The colormaps demonstrate that EP condensation occurs at nonzero momenta states. Indeed, EPs condense at the minima, where the photon group velocity turns into zero.
It can be seen that in both types of points, with one located at the M-point in $k$-space and the other located between $\Gamma$ and $K$ points (see Fig.~\ref{Fig2}), we observe a threshold-like behavior. 
At small, under-threshold pumping powers, the particles are thermally distributed at high energies above the ground state(s), as seen in Fig.~\ref{Fig3}a. The minima remain nearly unoccupied. With an increase of pumping power (Fig.~\ref{Fig3}b), the particles start to accumulate at the inflection points of the dispersion and we observe the bottleneck effect~\cite{bootleneck1, bootleneck2}. The last panel (Fig~\ref{Fig3}c) corresponds to the above-threshold pumping, when the particles start to Bose-condense at the minima.

Figure~\ref{Fig4} illustrates that the condensate formation varies for different points in $k$-space. We attribute this to the difference in the detunings of exciton energy and PC photon energy. Consequently, as expected, the less-detuned M point is more susceptible to condensation and exhibits a lower threshold.

%--------------------------------------------------------------

%{\it Conclusions.---} 
\section{Conclusions}
We have demonstrated the formation of organic exciton polaritons in a triangular lattice of AlN pillar, two-dimensional photonic crystal, and shown that Bose-Einstein condensation can take place at the minima of the band diagram where photon group velocity equals zero. 
Such dispersion acts as a set of traps for particles, and it can be employed to achieve polariton condensation at non-zero momenta, which may be useful, for example, in valleytronics~\cite{RefValley} and for spontaneous symmetry breaking. It should also be mentioned that one can replace our periodic (solid state) crystal with an optical lattice produced by crossed laser beams, as in cold atomic systems. Then it becomes easy to in situ vary the optical properties of the PC.

In the framework of our model, we found different particle densities at different points in $k$-space, controlled by the varying exciton-photon detuning at different points. 
In contrast to Bose-Einstein condensation in conventional quantum wells based on inorganic semiconductors, here organic materials with high molecular orientation provide selective coupling with TM (as opposed to TE) polarized modes and produce strong coupling due to a giant magnitude of the dipole moment, as opposed to regular inorganic excitons. 

\section*{Acknowledgement}
We thank T. Ellenbogen for the suggestion of this research project and useful discussions, and Joel Rasmussen (RECON) for a critical reading of our manuscript.
We acknowledge support of the IBS-R024-D1, the Australian Research Council Discovery Projects funding scheme (Project No. DE160100167), President of Russian Federation (Project No. MK-5903.2016.2), and Dynasty Foundation. D.V.K. thanks the IBS Center of Theoretical Physics of Complex Systems for hospitality.

%--------------------------------------------------------------

\end{document}